\documentclass[11pt]{article}
\usepackage{graphicx}
\usepackage{mathrsfs}
\usepackage{xypic}

\usepackage{color}
\usepackage{amsfonts}
\usepackage{enumitem}
\usepackage{xypic}
\usepackage{xy}
\usepackage{amsmath}
\usepackage{booktabs}
\usepackage{threeparttable}
\usepackage{amssymb}
\usepackage{bm}
\usepackage{CJK}
\usepackage{cite}
\usepackage{romanbar}
\newtheorem{theorem}{Theorem}[section]
\newtheorem{lemma}[theorem]{Lemma}
\newtheorem{definition}[theorem]{Definition}

\newtheorem{remark}[theorem]{Remark}
\newtheorem{proposition}[theorem]{Proposition}
\newtheorem{conjecture}[theorem]{Conjecture}

\numberwithin{equation}{section}

\newcommand{\ba}{\begin{array}}
	\newcommand{\ea}{\end{array}}
\newcommand{\bt}{\begin{tabular}}
	\newcommand{\et}{\end{tabular}}
\newcommand{\btb}{\begin{table}}
	\newcommand{\etb}{\end{table}}
\newcommand{\bc}{\begin{center}}
	\newcommand{\ec}{\end{center}}
\newcommand{\bea}{\begin{eqnarray}}
	\newcommand{\eea}{\end{eqnarray}}
\newcommand{\Bea}{\begin{eqnarray*}}
	\newcommand{\Eea}{\end{eqnarray*}}
\newcommand{\beq}{\begin{equation}}
	\newcommand{\eeq}{\end{equation}}

\newcommand{\spec}{\operatorname{Spec}}

\def \F {{\mathbb F}}

\begin{document}
	\baselineskip 16.5pt
	\title{A Geometric Approach to the Classification of Locally-APN Niho-Type Power Functions}
    \author{Chenxu Wang\footnote{Chenxu Wang is with the School of Mathematical Sciences, Capital Normal University, Beijing, China, 100048. Email: 2250501038@cnu.edu.cn},\and Jun Zhang\footnote{Jun Zhang is with the School of Mathematical Sciences, Capital Normal University, Beijing, China, 100048. Email: junz@cnu.edu.cn.}}

\date{}
\maketitle

\noindent\textbf{Abstract.}
Let \(q=2^m\) with \(m\geq2\). We develop a geometric approach to
the classification of locally-APN Niho-type power functions over
\(\mathbb F_{q^2}\). Our approach begins with a characterization of
\(\mathbb F_q\)-homogeneous functions: the conditions
\(D_F(a,F(a))=q\) and \(D_F(a,b)\leq2\) for every
\(a\in\mathbb F_{q^2}^*\) and \(b\neq F(a)\) hold if and only if
the projective graph of \(F\) is a \((q+1)\)-arc in
\(\mathrm{PG}(3,q)\). For Niho-type power functions, this differential
condition is equivalent to the locally-APN property. Applying the
classification of \((q+1)\)-arcs and the description of their
stabilizers, we prove, for \(m\geq3\), that
\(F_s(x)=x^{s(q-1)+1}\) is locally-APN if and only if
\[
s\equiv(2^k+1)^{-1}\pmod{q+1}
\]
for some \(1\leq k\leq2m-1\) satisfying
\(\gcd(k,m)=\gcd(2^k+1,q+1)=1\).
Together with the known case \(m=2\) and the sufficient condition
of Xie et al., this gives the complete classification and confirms
the conjecture of Cui, Luo, and Xiang.

\medskip


\section{Introduction}

The differential properties of functions over finite fields are of considerable significance in the analysis of substitution boxes used in symmetric cryptography. Since the advent of differential cryptanalysis, introduced by Biham and Shamir \cite{BihamShamir1991}, differential uniformity has served as a measure---introduced by Nyberg \cite{Nyberg1993}---of the largest multiplicity of an input-output difference pair. For a function
$F:\F_{2^n}\to\F_{2^n}$ and $a,b\in\F_{2^n}$, let
$$
S_F(a,b)=\{x\in\F_{2^n}:F(x+a)+F(x)=b\}.
$$
We call $D_F(a,b)=|S_F(a,b)|$ the entry of the difference distribution table of $F$ at $(a,b)$, and call
$$
\delta(F)=
\max_{\substack{a\in\F_{2^n}^*\\ b\in\F_{2^n}}}D_F(a,b)
$$
the differential uniformity of $F$.
Since solutions occur in pairs $\{x,x+a\}$ for $a\neq 0$, we have $\delta(F)\geq 2$. If $F$ attains this lower bound, i.e., $\delta(F)=2$, we call it an almost perfect nonlinear (APN) function.
 The construction and classification of APN functions have received considerable attention; see
\cite{CalderiniBudaghyanCarlet2021, Carlet2021,BlondeauNyberg2015, Pott2016,Budaghyan2014,BudaghyanCarletLeander2009,BrinkmannLeander2008,Calderini2020} for comprehensive
accounts of their cryptographic properties.

Power functions form a natural class in which to study
differential properties. For $F(x)=x^d$ and $a\ne0$,
the substitution $x=ay$ gives
$
D_F(a,b)=D_F(1,b/a^d).
$
Thus the differential distribution of a power function is determined by the row $D_F(1,b).$ 
To study functions whose differential multiplicities
are small outside a few exceptional values,
Blondeau, Canteaut, and Charpin
\cite{BlondeauCanteautCharpin2011} introduced
locally-APN power functions.  A power function
$F$ is called locally-APN if
$
\max_{b\in\F_{2^n}\setminus\F_2}D_F(1,b)=2.
$
This permits exceptional behavior at $b=0$ and $b=1$,
while retaining the APN bound at all other output
differences. Examples and their differential spectra
were studied in \cite{BlondeauCanteautCharpin2011}.
Further families, together with their boomerang spectra,
were investigated by Hu et al.
\cite{HuLiXuZengTang2023}.

 Let $m\geq 2$ and $q=2^m$. We consider the Niho-type power functions
$F_s(x)=x^{s(q-1)+1}$ over $\F_{q^2}$, where $s$ is taken modulo $q+1$.
Xie et al.~\cite{xie2022niho} proved that $F_s$ is locally-APN whenever
$s\equiv (2^k+1)^{-1}\pmod{q+1}$
for some positive integer $k$ satisfying
$\gcd(k,m)=\gcd(2^k+1,q+1)=1$. They also determined the differential
and boomerang spectra of this family. Moreover, their computations for
$2\leq m\leq 10$ indicated that, up to conjugation of the exponent and
inversion when the exponent is invertible modulo $q^2-1$, this family
accounts for all Niho-type locally-APN power functions over $\F_{q^2}$
\cite[Remark~1]{xie2022niho}. This observation raises a natural question:
does the family of Xie et al.\ exhaust all Niho-type locally-APN power
functions for every $m\geq 2$?

Cui, Luo, and Xiang~\cite{cui2026walsh} later arrived at the same
classification problem through their study of Walsh spectra. They
characterized Niho-type locally-APN power functions in terms of their
Walsh spectra and formulated the following conjecture.

\begin{conjecture}[{\cite[Conjecture~1]{cui2026walsh}}]\label{con}
Let $m\geq 2$ and $q=2^m$. Every Niho-type locally-APN power function
over $\F_{q^2}$ is of the form $F_s(x)=x^{s(q-1)+1}$ with
$
s\equiv 2^r(2^r+\varepsilon)^{-1}\pmod{q+1}
$
for some integer $r$ with $1\leq r<m$ and some
$\varepsilon\in\{-1,1\}$ satisfying
$\gcd(r,m)=\gcd(2^r+\varepsilon,q+1)=1$.
\end{conjecture}

The conjecture describes the family of Xie et al.\ in a different
parametrization. Indeed, since $2^m\equiv-1\pmod{q+1}$, we have
$$
\begin{aligned}
2^r(2^r-1)^{-1} &\equiv (2^{m-r}+1)^{-1}\pmod{q+1},\\
2^r(2^r+1)^{-1} &\equiv (2^{2m-r}+1)^{-1}\pmod{q+1},
\end{aligned}
$$
whenever the respective inverses exist. Thus the two choices of sign
correspond to $k=m-r$ and $k=2m-r$, respectively. Conversely, reducing
any admissible $k$ modulo $2m$ gives a representative in one of the
intervals $1\leq k<m$ and $m<k<2m$, and hence recovers one of the two
forms above.

The purpose of this paper is to develop a geometric approach to
the classification problem above. Our contributions are twofold.
First, we establish a general geometric characterization for
\(\mathbb F_q\)-homogeneous functions. We prove that
\[
D_F(a,F(a))=q,\qquad D_F(a,b)\leq2
\quad\text{for all }a\in\mathbb F_{q^2}^*
\text{ and }b\neq F(a),
\]
if and only if the projective graph of \(F\) is a
\((q+1)\)-arc in \(\mathrm{PG}(3,q)\). This result translates
a differential condition into a geometric property and applies
beyond the class of power functions.

Second, we apply this characterization to the Niho-type power
functions \(F_s(x)=x^{s(q-1)+1}\). The locally-APN property
implies the differential condition above, so the associated
projective graph is a \((q+1)\)-arc. We then use the classification
of these arcs by Casse and Glynn~\cite{casse1982solution} and the
description of their stabilizers by Ceria and Pavese~\cite{CeriaPavese2023}.
By constructing a linear map that preserves the associated arc
and comparing its eigenvalues with those of an element in the
stabilizer of a canonical arc, we derive the necessary restriction
on the exponent \(s\). Together with the sufficient condition
of Xie et al.~\cite{xie2022niho} and the known verification for
\(m=2\), this completes the classification and confirms
Conjecture~\ref{con}.

The rest of the paper is organized as follows.
Section~2 presents the geometric background and proves the
equivalence between the differential condition and the
\((q+1)\)-arc property.
Section~\ref{section3} applies this characterization to
Niho-type power functions and establishes the classification.
The final section summarizes the results and discusses
directions for further research.

\section{A Geometric Characterization of Homogeneous Functions}

In this section, we establish a geometric characterization
of a differential condition for $\F_q$-homogeneous functions.
We begin with  the
necessary background on projective geometry, and then
associate each such function with a projective point set.

Throughout this paper, let $q=2^m$ with $m\geq 2$, and let
$$
U_m=\{w\in\F_{q^2}^*:w^{q+1}=1\}.
$$
For $0\leq s\leq q$, put $F_s(x)=x^{s(q-1)+1}$.
We first introduce the Niho-type polynomials considered below.

\begin{definition}\label{def}
 A  Niho-type polynomial
over $\F_{q^2}$ is a polynomial of the form
$$
F(x)=xh(x^{q-1})
=\sum_{s=0}^{q}a_s F_s(x),$$
where $h(x)\in \F_{q^2}[x]$ and $\deg(h(x))\leq q.$ 
\end{definition}

\begin{remark}
 An $\mathbb{F}_q$-homogeneous function $F:\F_{q^2}\to \F_{q^2}$ is one satisfying $F(\lambda x)=\lambda F(x)$ for all $\lambda\in\mathbb{F}_q$ and $x\in \F_{q^2}$. Note that $F$ is $\mathbb{F}_q$-homogeneous if and only if $F(0)=0$ and $F(x)/x$ is constant on each multiplicative coset of $\mathbb{F}_q^*$ in $\F_{q^2}^*$. In fact, by \cite[Lemma~1]{wang2007cyclotomic}, an $\mathbb{F}_q$-homogeneous function is precisely a first-order cyclotomic mapping of index $q+1$, i.e., it admits the representation $F(x) = x h(x^{q-1})$. Consequently, a function $F:\F_{q^2}\rightarrow\F_{q^2}$ is  $\mathbb{F}_q$-homogeneous if and only if it admits a polynomial representation of the form given in Definition \ref{def}.
\end{remark}

Let \(\mathrm{PG}(3,q)\) denote the three-dimensional projective space over \(\mathbb{F}_q\). Throughout this paper, we represent its points in two coordinate systems: \([x_0 : x_1 : x_2 : x_3]\) with \(x_i \in \mathbb{F}_q\), and \([X : Y]_q\) with \(X, Y \in \mathbb{F}_{q^2}\). Fixing an \(\mathbb{F}_q\)-basis \(\{1, \theta\}\) of \(\F_{q^2}\), these two representations are identified via the map
\[
\varphi : [x_0 : x_1 : x_2 : x_3] \mapsto [x_0 + x_1\theta : x_2 + x_3\theta]_q.
\]
Here \([X : Y]_q\) denotes the one-dimensional \(\mathbb{F}_q\)-subspace spanned by \((X, Y) \neq (0, 0)\). Clearly, the map $\varphi$ is induced by an $\F_q$-linear isomorphism and therefore preserves collinearity and coplanarity.

We now introduce arcs in $\operatorname{PG}(3,q)$, which provide
the geometric framework for studying the locally-APN
property of the functions considered in this paper.

\begin{definition}
   A set $\mathcal{A}$ of $k\ge 4$ points in $\mathrm{PG}(3,q)$ is called a \emph{$k$-arc} if no four distinct points of $\mathcal{A}$ are coplanar, i.e., the representative vectors of any four distinct points of $\mathcal{A}$ are linearly independent over $\mathbb{F}_q$.
\end{definition}

Casse and Glynn \cite{casse1982solution} completely classified the \((q+1)\)-arcs of \(\operatorname{PG}(3,q)\) when \(q\) is a power of \(2\). We recall this classification, together with the description of their full stabilizers in \(\operatorname{PGL}(4,q)\). These results play a central role in our approach, allowing us to translate the geometric structure of the associated arcs into restrictions on the Niho exponents.

\begin{proposition}[\cite{casse1982solution}]
\label{prop:arcclass}
Let $q=2^m$ with $m\geq3$, and let $\mathcal A$ be a
$(q+1)$-arc of $\mathrm{PG}(3,q)$.
Then there exist an integer $e$ with $1\leq e<m$ and
$\gcd(e,m)=1$, and an element $P\in\mathrm{PGL}(4,q)$,
such that $P(\mathcal A)=C_e$, where
$$
C_e=
\{[1:z:z^{2^e}:z^{2^e+1}]:z\in\F_q\}
\cup\{[0:0:0:1]\}.
$$
\end{proposition}

\begin{proposition}[\cite{CeriaPavese2023}]
\label{prop:arcstab}
Let $q=2^m$ with $m\geq3$, and let $1\leq e<m$ satisfy
$\gcd(e,m)=1$.
Define
$$
\rho_e:\operatorname{SL}(2,q)\longrightarrow
\operatorname{GL}(4,q),
\qquad
H\longmapsto H^{(2^e)}\otimes H,
$$
where $H^{(2^e)}$ is obtained by raising each entry of $H$
to the $2^e$-th power, and $\otimes$ denotes the
Kronecker product.
Then the induced homomorphism
$$
\bar\rho_e:\operatorname{SL}(2,q)\longrightarrow
\operatorname{PGL}(4,q),
\qquad
H\longmapsto[\rho_e(H)]
$$
is injective, and
$$
\operatorname{Stab}_{\operatorname{PGL}(4,q)}(C_e)
=
\bar\rho_e(\operatorname{SL}(2,q)).
$$
Moreover, $\det\rho_e(H)=1$ for every
$H\in\operatorname{SL}(2,q)$.
\end{proposition}

Next, we associate to each \(\mathbb{F}_q\)-homogeneous function a projective point set, thereby translating the function into a geometric object in \(\operatorname{PG}(3,q)\).

\begin{definition}
    Let \(F:\F_{q^2}\rightarrow \F_{q^2}\) be an \(\mathbb{F}_q\)-homogeneous function. We define the projective graph \(\mathcal{A}_F\) of \(F\) as the subset of \(\mathrm{PG}(3,q)\) given by
\[
\mathcal{A}_F := \{[x : F(x)]_q : x \in \mathbb{F}_{q^2}^*\}.
\]
\end{definition}

\begin{remark}\label{rm1}
Using homogeneity, we have \([x : F(x)]_q = [\lambda x : F(\lambda x)]_q\) for each \(\lambda \in \mathbb{F}_q^*\). Thus, the map \(x \mapsto [x : F(x)]_q\) is constant on each \(\mathbb{F}_q^*\)-orbit in \(\mathbb{F}_{q^2}^*\). Since \(U_m\) is precisely a set of representatives for these orbits, we obtain
\[
\mathcal{A}_F = \{[w : F(w)]_q : w \in U_m\},
\]
and \(|\mathcal{A}_F| = q + 1\).
\end{remark}

Having identified $\mathcal{A}_F$ as a set of $q+1$ points, we now establish an equivalence between the arc property of $\mathcal{A}_F$ and the differential properties of the function $F$. The following theorem shows that this geometric property is equivalent to a purely differential condition.

\begin{theorem}\label{thm:Locally-arc}
Let $q=2^m$ with $m\geq 2$, and let $F : \F_{q^2} \to \F_{q^2}$ be an \(\mathbb{F}_q\)-homogeneous function. The following conditions are equivalent.
\begin{enumerate}
    \item[(i)] For every \(a \in \F_{q^2}^*\),
$D_F(a, F(a)) = q,$ and for $b\ne F(a),$ $ D_F(a, b)\leq 2.$ 
\item[(ii)] \(\mathcal{A}_F\) is a \((q+1)\)-arc of \(\mathrm{PG}(3, q)\).
\end{enumerate}
\end{theorem}

\begin{remark}\label{rm2}
By the \(\mathbb{F}_q\)-homogeneity of \(F\), every \(x \in a\mathbb{F}_q\) satisfies \(F(x+a)+F(x)=F(a)\). Thus, the equality \(D_F(a,F(a))=q\) in condition~{\rm(i)} means that
$S_F(a,F(a)) = a\mathbb{F}_q.$
\end{remark}
\begin{proof}
\((i)\Longrightarrow(ii).\)
Suppose that condition~{\rm(i)} holds. Remark~\ref{rm1} gives
$|\mathcal A_F|=q+1$. It suffices to show that $\mathcal A_F$ is an arc.

We first show that no three distinct points of $\mathcal A_F$ are
collinear. Suppose otherwise. By the $\F_q$-homogeneity of $F$, we
may choose $y_1,y_2,y_3\in\F_{q^2}^*$ from distinct $\F_q^*$-cosets
such that
$$
y_1+y_2+y_3=0,~ F(y_1)+F(y_2)+F(y_3)=0.
$$
It follows that
$y_1\in S_F(y_2,F(y_2))=y_2\F_q$ by Remark~\ref{rm2},
contradicting the choice of $y_1$ and $y_2$.

Now suppose that four distinct points of $\mathcal A_F$ are coplanar.
Since no three are collinear, all four coefficients in a nontrivial
linear relation among their representatives are nonzero. By homogeneity,
we may therefore choose $y_1,y_2,y_3,y_4\in\F_{q^2}^*$ from distinct
$\F_q^*$-cosets such that
$$
\sum_{i=1}^4 y_i=0,\qquad \sum_{i=1}^4 F(y_i)=0.
$$
Set
$$
a=y_1+y_2=y_3+y_4,\qquad
b=F(y_1)+F(y_2)=F(y_3)+F(y_4).
$$
Then $a\ne0$ and $y_1,y_2,y_3,y_4\in S_F(a,b)$.
If $b\ne F(a)$, this contradicts $D_F(a,b)\leq2$.
If $b=F(a)$, then Remark~\ref{rm2} gives
$y_1,y_2\in a\F_q$, again contradicting the fact that $y_1$ and
$y_2$ belong to distinct $\F_q^*$-cosets.
Hence no four distinct points of $\mathcal A_F$ are coplanar,
and $\mathcal A_F$ is a $(q+1)$-arc.

\((ii)\Longrightarrow(i).\)
First we prove that $D_F(a,F(a))=q$ for each $a\ne0$.
By homogeneity, $a\F_q\subseteq S_F(a,F(a))$.
If there were a solution $x\notin a\F_q$, then
$[x:F(x)]_q$, $[x+a:F(x+a)]_q$, and $[a:F(a)]_q$ would be
three distinct points of $\mathcal A_F$. The vector identity
$$
(x,F(x))+(x+a,F(x+a))=(a,F(a))
$$
shows that these points are collinear, contradicting the arc property.
Thus $S_F(a,F(a))=a\F_q$ and $D_F(a,F(a))=q$.

Next, fix $a\ne0$ and $b\ne F(a)$. Suppose, to the contrary, that
$D_F(a,b)>2$. Since solutions occur in pairs, there exist four
distinct elements $x,x+a,y,y+a$ such that
$$
F(x)+F(x+a)=F(y)+F(y+a)=b.
$$
Since $b\ne F(a),$ we obtain that none of these elements lies in $a\F_q.$
Write $P_z=[z:F(z)]_q$ and $Q=[a:b]_q$. By homogeneity and
$b\ne F(a)$, we have $Q\notin\mathcal A_F$. The vector identities
$$
(x,F(x))+(x+a,F(x+a))
=(a,b)
=(y,F(y))+(y+a,F(y+a))
$$
show that both lines $P_xP_{x+a}$ and $P_yP_{y+a}$ pass through $Q$.

We claim that $xy^{-1}\notin\F_q$. Suppose otherwise, and write
$x=\lambda y$ with $\lambda\in\F_q^*\setminus\{1\}$. Then $P_x=P_y$.
On the other hand, $P_{x+a}\ne P_{y+a}$: indeed, if
$x+a=\mu(y+a)$ for some $\mu\in\F_q^*$, then
$(\lambda-\mu)y=(\mu-1)a$. Since $y\notin a\F_q$, this forces
$\lambda=\mu=1$, a contradiction. Thus $P_x$, $P_{x+a}$, and
$P_{y+a}$ are three distinct graph points on the line $P_xQ$,
contradicting the arc property. This proves the claim.

Interchanging $x$ with $x+a$ or $y$ with $y+a$ gives the same
conclusion for the other three cross-pairs. Hence $P_x$, $P_{x+a}$,
$P_y$, and $P_{y+a}$ are four distinct points of $\mathcal A_F$.
Since the lines $P_xP_{x+a}$ and $P_yP_{y+a}$ meet at $Q$, these
four points are coplanar, again contradicting the arc property.
Therefore $D_F(a,b)\leq2$.
\hfill$\Box$
\end{proof}

\section{Complete Characterization of Locally-APN Niho-Type Power Functions}\label{section3}
In this section, we apply the geometric characterization
established above to the power functions
$F_s(x)=x^{s(q-1)+1}$ over $\F_{q^2}$, where $q=2^m$
and $m\geq2$.
Our goal is to determine all parameters $s$ for which
$F_s$ is locally-APN.
We first recall the sufficient conditions obtained
by Xie et al.~\cite{xie2022niho}.

\begin{theorem}[{\cite{xie2022niho}}]\label{thm:s}
Let $m\geq2$ and $k\geq1$ be integers satisfying
$\gcd(k,m)=\gcd(2^k+1,2^m+1)=1$, and let $s$ denote the
multiplicative inverse of $2^k+1$ modulo $2^m+1$.
Then $F_s(x)=x^{s(2^m-1)+1}$ is locally-APN over $\F_{2^{2m}}$.
\end{theorem}

Conversely, using the connection between the locally-APN
property and arcs, we show that every Niho type locally-APN
power function belongs to the family described in
\cite{xie2022niho}, thereby completing the classification. In view of the computational verification for $m=2$ reported in
\cite[Remark~1]{xie2022niho}, it remains to prove the necessity
of the condition for $m\geq3$.  More precisely, we obtain the following result.
\begin{theorem}\label{thm:n}
    Let $m\geq 3.$ If \(F_s\) is locally-APN, then there is an integer \(1\le k\le 2m-1\) such that $s\equiv (2^k+1)^{-1} \pmod{q+1}$ and $\gcd(k,m)=\gcd(2^k+1,q+1)=1.$
\end{theorem}

We first recall two consequences of the locally-APN
property.
They will allow us to apply
Theorem~\ref{thm:Locally-arc} to $F_s$ and hence
associate a $(q+1)$-arc with this power function.
\begin{lemma}[\cite{cui2026walsh,xie2022niho} ]\label{lem3}
    If $F_s(x)$ is  locally-APN,  the following hold.
    
    (i) $\gcd(s,q+1)=\gcd(s-1,q+1)=1.$ 

    (ii) $S_{F_s}(1,1)=\F_q$ and $D_{F_s}(1,b)\leq 2$ for $b\ne 1.$ 
\end{lemma}

For completeness, the differential condition in
Theorem~\ref{thm:Locally-arc} is equivalent to the locally-APN
property for the power functions considered here. Indeed, Lemma~\ref{lem3}
gives one implication; the scaling identity for power functions extends
the corresponding statements from \(a=1\) to every \(a\neq0\).
Conversely, suppose that
\(D_{F_s}(1,1)=q\) and \(D_{F_s}(1,b)\leq2\) for every \(b\neq1\).
Since
\[
\sum_{b\in\mathbb F_{q^2}}D_{F_s}(1,b)=q^2
\quad\text{and}\quad
D_{F_s}(1,0)\leq2,
\]
we have
\[
\sum_{b\in\mathbb F_{q^2}\setminus\mathbb F_2}
D_{F_s}(1,b)
=q^2-q-D_{F_s}(1,0)>0,
\]
where \(q\geq4\). Each differential multiplicity is even,
and each term in this sum is at most \(2\). Hence at least one
term equals \(2\), so \(F_s\) is locally-APN.

The proof also requires the eigenvalues and determinants
of multiplication maps on $\F_{q^2}$, viewed as linear
transformations over $\F_q$.
We record these elementary facts below.

\begin{lemma}[\cite{LidlNiederreiter1997,LiesenMehrmann2015}]\label{lem:eig and det}
Let $q=p^m$ be a prime power, and write
$N(a)=a^{q+1}$ for the norm from $\F_{q^2}$ to $\F_q$.
All eigenvalues below are taken over an algebraic closure of
$\F_q$ and counted with algebraic multiplicity.
\begin{enumerate}
    \item For $a\in\F_{q^2}$, the $\F_q$-linear map $x\mapsto ax$ on $\F_{q^2}$ has characteristic polynomial $(X-a)(X-a^q)$ and determinant $N(a)$. Moreover, for $a,b\in\F_{q^2}$, the $\F_q$-linear map $T:(x,y)\mapsto(ax,by)$ on $\F_{q^2}^2$ has spectrum
$
\spec(T)=\{a,a^q,b,b^q\}
$
and determinant $\det T=N(ab)$.
   
 \item Let $A\in M_u(\F_q)$ and $B\in M_v(\F_q)$, with
$\spec(A)=\{\alpha_1,\ldots,\alpha_u\}$ and $\spec(B)=\{\beta_1,\ldots,\beta_v\},$
where spectra are regarded as multisets over an algebraic
closure of $\F_q$. Then
\[
\spec(A\otimes B)
=\{\alpha_i\beta_j:1\leq i\leq u,\ 1\leq j\leq v\},
\]
and
$\det(A\otimes B)=(\det A)^v(\det B)^u.$
\end{enumerate}
\end{lemma}

We now give the proof of Theorem~\ref{thm:n}.

\noindent \textbf{Proof of Theorem \ref{thm:n}.}
   Let \(\zeta\) generate \(U_m\), and put \(r = 1 - 2s\). Since \(w^{q-1} = w^{-2}\) for \(w \in U_m\), we have \(F_s(w) = w^r\). Define
\[
\mathcal{A}_r := \mathcal{A}_{F_s} = \{[w : w^r]_q : w \in U_m\}.
\]
Consider the \(\mathbb{F}_q\)-linear map
\begin{align*}
    T:\F_{q^2}^2&\longrightarrow \F_{q^2}^2,\\
    (x,y)&\longmapsto (\zeta x,\zeta^r y).
\end{align*}
On the one hand, since $\zeta^q=\zeta^{-1}$,
Lemma~\ref{lem:eig and det} gives
$
\det T=N(\zeta^{r+1})=1
$
and
\begin{equation}\label{specT}
    \spec(T)=\{\zeta,\zeta^{-1},\zeta^r,\zeta^{-r}\}.
\end{equation}

By Lemma~\ref{lem3}(ii) and the scaling identity
$D_{F_s}(a,b)=D_{F_s}(1,b/F_s(a))$ for $a\ne0$, condition~{\rm(i)}
of Theorem~\ref{thm:Locally-arc} holds. Hence $\mathcal A_{F_s}$ is a $(q+1)$-arc.
By Proposition~\ref{prop:arcclass}, there exist an integer $e$
with $1\le e<m$ and $\gcd(e,m)=1$, and a matrix
$P\in\operatorname{GL}(4,q)$ such that $[P](\mathcal A_r)=C_e$.
For every $w\in U_m$, we have
$$
[T]([w:w^r]_q)=[\zeta w:(\zeta w)^r]_q.
$$
Since multiplication by $\zeta$ permutes $U_m$, it follows that
$[T]$ stabilizes $\mathcal A_r$. Hence
$$
[PTP^{-1}]\in\operatorname{Stab}_{\operatorname{PGL}(4,q)}(C_e).
$$
Applying Proposition~\ref{prop:arcstab}, we obtain
$
PTP^{-1}=c\rho_e(H)
$
for some $c\in\F_q^*$ and $H\in\operatorname{SL}(2,q)$.
Since $\det T=\det\rho_e(H)=1$, taking determinants gives
$c^4=1$. As $\F_q$ has characteristic two, this forces $c=1$.
Therefore,
\begin{equation}\label{eq:PTP}
    PTP^{-1}=\rho_e(H)=H^{(2^e)}\otimes H.
\end{equation}

Moreover, since $\zeta$ generates the cyclic group $U_m$
of order $q+1$, the map $T$ has order $q+1$.
Thus $\rho_e(H)$ also has order $q+1$.
Since $\rho_e$ is an injective group homomorphism,
$$
\rho_e(H)^n=\rho_e(H^n)=I_4
\quad\Longleftrightarrow\quad H^n=I_2.
$$
Consequently, $\operatorname{ord}(H)=q+1$.

The minimal polynomial of $H$ divides $X^{q+1}-1$,
which splits into distinct linear factors over $\F_{q^2}$.
Hence $H$ is diagonalizable over $\F_{q^2}$.
Since $\det H=1$, its eigenvalues can be written as
$\eta$ and $\eta^{-1}$.
It follows that
$
\operatorname{ord}(\eta)=\operatorname{ord}(H)=q+1.
$
Thus $\eta$ generates $U_m$, and we may write
$\eta=\zeta^j$ with $\gcd(j,q+1)=1$. Note that the eigenvalues of $H^{(2^e)}$ are $\eta^{2^e}$ and $\eta^{-2^e}.$
By Lemma~\ref{lem:eig and det}, we obtain
\begin{equation}\label{spec2}
\spec(\rho_e(H))
=
\{\zeta^{j(2^e+1)},\zeta^{j(2^e-1)},
  \zeta^{-j(2^e-1)},\zeta^{-j(2^e+1)}\}.
\end{equation}

Comparing \eqref{specT} and \eqref{spec2}, and using that
\(\zeta\) has order \(q+1\), we obtain the multiset congruence
\[
\{1,-1,r,-r\}
=
\{j(2^e+1),-j(2^e+1),j(2^e-1),-j(2^e-1)\}
\pmod{q+1}.
\]
Put \(n=q+1\), \(A=2^e+1\), and \(B=2^e-1\).
The four residues \(A,-A,B,-B\) are pairwise distinct modulo \(n\).
Indeed, \(0<B<A<n\), \(n\) is odd, and
\(A-B=2\) and \(A+B=2^{e+1}<n\).
Since \(j\) is invertible modulo \(n\), the same is true of
\(jA,-jA,jB,-jB\). We distinguish two cases according to
which of the pairs \(\{\pm jA\}\) and \(\{\pm jB\}\) contains \(1\).
All inverses below are taken modulo \(n\).

\textbf{Case 1:} \(1\equiv\pm jA\pmod n\).
Then \(A\) is invertible modulo \(n\), and the multiset congruence
gives \(r\equiv\pm BA^{-1}\pmod n\). Since
\(r=1-2s\), we obtain
\[
s\equiv A^{-1}
\quad\text{or}\quad
s\equiv 2^eA^{-1}=(1+2^{-e})^{-1}
\pmod n.
\]
The first possibility corresponds to \(k=e\), while the second
corresponds to \(k=2m-e\), because
\[
2^{2m-e}+1\equiv2^{-e}(2^e+1)\pmod n.
\]

\textbf{Case 2:} \(1\equiv\pm jB\pmod n\).
Then \(B\) is invertible modulo \(n\), and
\(r\equiv\pm AB^{-1}\pmod n\). Consequently,
\[
s\equiv(1-2^e)^{-1}
\quad\text{or}\quad
s\equiv(1-2^{-e})^{-1}
\pmod n.
\]
Since \(2^m\equiv-1\pmod n\), these two possibilities correspond
to \(k=m+e\) and \(k=m-e\), respectively, because
\[
2^{m+e}+1\equiv1-2^e\pmod n,
\qquad
2^{m-e}+1\equiv1-2^{-e}\pmod n.
\]

In all four cases, \(1\leq k\leq2m-1\) and
\(k\equiv\pm e\pmod m\). Hence
\(\gcd(k,m)=\gcd(e,m)=1\). Moreover,
\(2^k+1\) is a unit multiple modulo \(n\) of \(A\) in Case~1
and of \(B\) in Case~2. Thus
\(\gcd(2^k+1,q+1)=1\), completing the proof.
\hfill \(\square\)

\section{Conclusion}

In this paper, we developed a geometric approach to the
classification of locally-APN Niho-type power functions.
Our first main result establishes a general correspondence
for \(\mathbb F_q\)-homogeneous functions: the differential
conditions
\[
D_F(a,F(a))=q,\qquad
D_F(a,b)\leq2\quad\text{for }b\neq F(a),
\]
for every \(a\in\mathbb F_{q^2}^*\), hold if and only if
the projective graph of \(F\) is a \((q+1)\)-arc in
\(\mathrm{PG}(3,q)\). This characterization provides a
geometric way to study differential properties of
homogeneous functions.

We then applied this correspondence to Niho-type power
functions. Using the classification of \((q+1)\)-arcs
and the description of their stabilizers, we derived the
necessary condition on the exponent. Combining this
condition with the sufficient condition of Xie et al.
completes the classification for \(m\geq3\); the case
\(m=2\) is covered by the verification in
\cite[Remark~1]{xie2022niho}. Thus our result confirms
Conjecture~\ref{con}.

It would be interesting to extend this geometric approach
to broader classes of \(\mathbb F_q\)-homogeneous functions,
including Niho-type binomials and other sparse polynomials.

\bibliographystyle{plain}	
\bibliography{References}
\end{document}